\documentclass{iau} 
\usepackage{graphicx}

\title[ELTs for complex populations around the GC] 
{Extremely large telescopes \\ 
for complex stellar populations \\
around the Galactic centre}

\author[Noriyuki Matsunaga]   
{Noriyuki Matsunaga}

\affiliation{Department of Astronomy, The University of Tokyo, \\ 7-3-1 Hongo, Bunkyo-ku, Tokyo 113-0033, Japan
\\ email: {\tt matsunaga@astron.s.u-tokyo.ac.jp} } 

\pubyear{2018}
\volume{347}  
\jname{Early Science with ELTs (EASE)}
\editors{N. Przybilla, K. Pollard \& A. Calamida, eds.}
\begin{document}

\maketitle

\begin{abstract}
The Galactic centre and its surrounding space
are important in studying galaxy-scale evolution,  
and stellar populations therein are expected to have imprints of 
the long-term evolution.
Interstellar extinction, however, severely limits optical
observations, thereby requiring infrared observations.
In addition, many systems from those in the proximity of
the central black hole to foreground objects in the disc
overlap in each sightline, which complicates the interpretation
of observations of a wide variety of objects.
We discuss some important issues concerning
the central regions, particularly
the Galactic bulge and the Nuclear Stellar Disc
(also known as the Central Molecular Zone).
An obvious advantage of Extremely Large Telescopes (ELTs)
is the deeper limiting magnitudes,
but we emphasise the importance of the synergy between
the data of deep ELTs and other observational data 
(e.g.\  astrometric measurements and the detection of interstellar absorption lines) in order to
disentangle the complex stellar populations.
\end{abstract}

\firstsection 
\section{Introduction}
\label{sec:Intro}

The region within a few kiloparsecs from the Galactic centre (GC, hereinafter)
is extremely interesting and complicated.
In the very centre, a super massive black hole (BH)
of a few million solar masses is surrounded by
the Nuclear Stellar Cluster (NSC),
which hosts stars---more than $10^7~{\rm M}_\odot$
within a few parsecs (\cite[Genzel \etal\ 2010]{Genzel-2010}).
Up to {$\sim$}250~pc around the centre is a disc system
composed of stars, ${\sim}10^9~{\rm M}_\odot$,
and interstellar gas (mainly molecular) and dust;
the system is called the Nuclear Stellar Disc (NSD)
and the Central Molecular Zone (CMZ)
(\cite[Morris \& Serabyn, 1996]{Morris-1996}; \cite[Launhardt \etal\ 2002]{Launhardt-2002}).
Around the NSD/CMZ and extended out to a few kiloparsecs
is the Galactic bulge (GB),
which is made up of a large number of stars,
${\sim}2\times 10^{10}~{\rm M}_\odot$ (\cite[Valenti \etal\ 2016]{Valenti-2016}).
Approximately at the Galactocentric distance of 3~kpc,
the Galactic disc takes over as the main contributor
to stellar populations in the Galactic disc. Generally speaking,
observational studies on the innermost part of the disc 
are lacking, and the stellar populations
around the interface between the GB and the disc remain elusive.

These regions, at around the distance of {$\sim$}8.3~kpc
(\cite[de Grijs \& Bono, 2016]{deGrijs-2016}),
provide us with unique opportunities to investigate in detail
the central components
of a galaxy by observing individual stars and interstellar matter (ISM)
down to a small scale.
In fact, the last decade has seen many new results concerning
stellar populations therein based on 
rapidly growing observational data from many large-scale, 
both photometric and spectroscopic, surveys targeting the GB
(see the review by \cite[Barbuy \etal\ 2018]{Barbuy-2018}).
There are, however, a few difficulties in studying the central regions.
First, strong interstellar extinction prevents us from observing 
the stars in shorter wavelengths
and measuring their intrinsic brightness and colours.
Dealing with the extinction towards the GC is especially hard,
because
it is patchy with respect to both the celestial position and
the ling-of-sight (LoS) distance.
Second, the interpretation of observational data is often complicated by
the fact that various components of stars and ISM,
including those in the foreground and background disc,
are mixed along the LoS.
In Section~\ref{sec:Review}, we describe these problems
as well as review some recent results and
important unsolved issues.
We then discuss some approaches in Section~\ref{sec:Approach},
which will become more and more practical
by using data from Extremely Large Telescopes (ELTs)
and other facilities.
The high sensitivity of ELTs is a simple and clear advantage
in observing faint stars (or stars of a given brightness with
significantly shorter observational times and/or higher precision),
as we present
in Section~\ref{sec:ELTs}, but combining various types of data
is crucial for disentangling the mixed stellar populations around the GC.

\section{Recent progress and unsolved matters}
\label{sec:Review}

\subsection{Nuclear Stellar Cluster (NSC)}
\label{sec:NSC}

With intensive and sustained efforts, many publications
on the NSC have been reported in the literature. 
The results based on near-infrared data
cover a broad range including
the mass of the central BH, the distance to the GC
based on stellar orbits around Sgr~A$^*$,
and stellar populations within the cluster
(\cite[Genzel \etal\ 2010]{Genzel-2010}; \cite[Fritz \etal\ 2016]{Fritz-2016},
and references therein).
Detailed observations of stars and ISM in the proximity of
the central BH allow us to investigate exciting events, which are
useful for testing fundamental processes in astronomy and physics
(\cite[Plewa \etal\ 2017]{Plewa-2017}; \cite[GRAVITY Collaboration, 2018]{GRAVITY-2018}).
This cluster is also unique among stellar clusters in the Galaxy because
it hosts stellar populations with a wide range of ages from a few megayears 
to older than 5~Gyr (\cite[Pfuhl \etal\ 2011]{Pfuhl-2011}; see also a report of RR Lyrae in \cite[Dong \etal\ 2017]{Dong-2017}).
A long-standing issue is the formation and growth of the NSC
and the presence of young stars (e.g.\  \cite[Antonini, 2013]{Antonini-2013}; \cite[Lu \etal\ 2014]{Lu-2014}; \cite[Yelda \etal\ 2014]{Yelda-2014});
are they formed {\it in situ}, and if so where does the ISM come from
and how can
it lose its angular momentum and fall into the central part?
The most likely source for such ISM is the gas of the CMZ,
but the mechanism(s) of the gas transfer is/are unclear
(\cite[Gallego \& Cuadra, 2017]{Gallego-2017}).
Although the main focus of this contribution is 
on the NSD/CMZ and the GB, and not on the NSC, 
studying the evolution of ISM and stars in the NSD/CMZ
is relevant to important questions about the NSC
and the related evolution of the Galaxy.
For example, the circulation of ISM among these systems may be
important for the growth of the central BH
and the BH--host coevolution (\cite[Kormendy \& Ho, 2013]{Kormendy-2013}).

\subsection{Nuclear Stellar Disc/Central Molecular Zone (NSD/CMZ)}
\label{sec:NSDCMZ}

The presence of the high-density disc-like system has been
recognised for a long time, since early infrared and radio
observations (\cite[Blitz \etal\  1993]{Blitz-1993}; \cite[Morris \& Serabyn, 1996]{Morris-1996}),
and the overall and basic properties 
of the NSD/CMZ were summarised by \cite{Launhardt-2002}.
This relatively small region harbours 3--10~\% of
the total molecular gas and star formation
of the Galaxy (\cite[Kauffmann, 2017]{Kauffmann-2017a}).
There exist various types of stars of a wide range of ages, including
young stars such as blue luminous variables and Wolf--Rayet stars
(especially those in the Arches and Quintuplet clusters aged at {$\sim$}4~Myr;
\cite[Stolte \etal\ 2014]{Stolte-2014}, and references therein) and
a few classical Cepheids aged {$\sim$}25~Myr
(Matsunaga \etal\ \cite[2011]{Matsunaga-2011}, \cite[2015]{Matsunaga-2015}).
Besides, the observations of ISM, mainly in radio wavelengths, have been
very active (\cite[Morris \& Serabyn, 1996]{Morris-1996}; \cite[Kauffmann \etal\ 2017]{Kauffmann-2017b}).
Here, we raise two questions concerning star formation and
stellar ages in the NSD/CMZ. 

{\underline{\it How to supply gas for the episodic star formation in the NSD/CMZ}}.
\\
The star formation rate in the NSD/CMZ in the past few megayears
has been estimated at ${\sim}0.1~M_\odot\,{\rm yr}^{-1}$ (\cite[Barnes \etal\ 2017]{Barnes-2017}).
The ISM must be supplied from somewhere outside to explain
the stellar mass
(${\sim}10^9~M_\odot$) as well as the ISM mass (${\sim}10^7~M_\odot$)
in the NSD/CMZ (\cite[Launhardt \etal\ 2002]{Launhardt-2002}).
One plausible path of the gas transfer is
the highly elongated orbits along the bar structure
through which the ISM in the inner disc can fall to the CMZ 
(\cite[Binney \etal\ 1991]{Binney-1991}; \cite[Kim \etal\ 2011]{Kim-2011}; \cite[Sormani \etal\ 2015]{Sormani-2015}).
Moreover, it has been suggested that the star formation in the NSD/CMZ 
is episodic, with a time scale of order 20~Myr
(\cite[Stark \etal\ 2004]{Stark-2004}; \cite[Krumholz \& Kruijssen, 2015]{Krumholz-2015}; \cite[Krumholz \etal\ 2017]{Krumholz-2017}).
In fact, based on the period distribution of Cepheids in the NSD,
\cite{Matsunaga-2011} concluded that the star formation
rate 30--70~Myr ago was lower than the rate 20--30~Myr ago.
In addition to the three classical Cepheids at {$\sim$}25~Myr
found in the original survey, \cite{Matsunaga-2015} found
a new classical Cepheid which is very similar in age and other properties
to the three, and
also considered to be within the NSD. 
Together, the four Cepheids indicate a star formation rate
of $0.1^{+0.2}_{-0.05}~{\rm M}_\odot\,{\rm yr}^{-1}$ in 20--30~Myr ago,
which is higher than the rate in 30--70~Myr ($< 0.02~{\rm M}_\odot\,{\rm yr}^{-1}$, 
as inferred from the absence of Cepheids with such ages).
To better understand the physical conditions of the star formation
in the NSD/CMZ,
the distribution and substructures of the ISM and young stars within this region is important.
It has been long known that the distribution of the CMZ ISM is 
highly asymmetric in Galactic longitude and radial velocity, $v_{\rm LoS}$
(\cite[Bally \etal\ 1988]{Bally-1988}). However, it is difficult to know
the LoS positions of the ISM, as well as those of stars within the NSD/CMZ.
The information on $v_{\rm LoS}$ is useful;
\cite{Molinari-2011}, for example, discovered an outstanding
$\infty$-shaped feature, in projection,
of dust and combined the dust map with
$v_{\rm LoS}$ data of CS molecules to infer the twisted elliptical ring
(see also \cite[Kruijssen \etal\ 2015]{Kruijssen-2015}).
With many components mixed along the LoS,
however, ambiguity tends to remain in such analysis.
Whereas an overall distribution characterised by a ring
(not a pan cake) structure has been seen in many studies,
recent studies suggest that various substructures such as
small spiral arms (\cite[Sofue, 1995]{Sofue-1995}; \cite[Sawada \etal\ 2004]{Sawada-2004}) and streams (\cite[Kruijssen \etal\ 2015]{Kruijssen-2015})
can be naturally produced in hydrodynamical simulations
of barred galaxies, where the gas transfer from
the bar to the central part is also predicted
(Sormani \etal\ \cite[2015]{Sormani-2015}, \cite[2018]{Sormani-2018}; \cite[Ridley \etal\ 2017]{Ridley-2017}).
Detailed mappings
of the ISM and stellar populations making use of astrometric data,
which give proper motions, and interstellar absorption features
would be useful to verify the distribution of the ISM.
Moreover, knowing the locations and kinematics of stars,
especially young stars, within the NSD/CMZ is also crucial
for understanding the evolution of this system.

{\underline{\it Oldest stars in the NSD/CMZ and the formation of the system}}.
\\
Whereas the NSD/CMZ clearly hosts young stars as mentioned above,
the age of the oldest stars therein is unclear. Because many old stars
exist in the GB, it is necessary to identify stars in the NSD/CMZ
based on kinematics. 
A group of evolved AGB stars with maser emission, in particular
OH/IR stars, are known to show a concentration towards the NSD/CMZ
and kinematics consistent with its disc rotation 
(\cite[Lindqvist \etal\ 1992]{Lindqvist-1992}; \cite[Deguchi \etal\ 2004]{Deguchi-2004}).
The ages of these objects are not clear, but are considered to
be intermediate, say 0.1--5~Gyr.
SiO masers, at the GC distance, are mostly detected in
Miras with period longer than {$\sim$}400~days
(\cite[Deguchi \etal\ 2004]{Deguchi-2004}), which
are supposedly younger than those in globular clusters
that have $P\lesssim 300$~days
(see, e.g.\ \cite[Catchpole \etal\ 2016]{Catchpole-2016}, concerning ages and periods of Miras in the bulge).
In contrast, RR Lyrae and type II Cepheids, if found in the NSD/CMZ, would be
conclusive tracers of an old population. However, they are rather faint
owing to the strong extinction, $2\lesssim A_K \lesssim 3$
(e.g.\  see Fig.~1 in
\cite[Contreras Ramos \etal\ 2018]{ContrerasRamos-2018}),
especially when we need to know
their kinematics to distinguish them from the GB population.
\cite{Dong-2017} reported the discovery of RR Lyrae in
the NSC using deep HST/WFC3 images, but the oldest stars in the NSD/CMZ
are not necessarily coeval with the counterparts in the NSC.
\cite{Matsunaga-2013} suggested the presence of
old stars in the NSD/CMZ based on the excess of
short-period ($<350$~days) Miras
and type II Cepheids over the density gradient of these objects in the GB,
but this result is indirect and also needs to be confirmed with
more complete samples of these variable stars.
Thus, the oldest stars in the NSD/CMZ remain to be identified.
It would also be sensible to ask if the first stars of the system 
were formed therein or accumulated through a dynamical process.
This may be related to another question, `when did the bar form?',
if we accept the gas transfer through the bar as the ISM fuelling mechanism
for the CMZ and that the first stars were formed {\it in situ}.
To answer these questions, deeper surveys of variable stars and
the kinematic identification of the NSD/CMZ members will be useful.

\subsection{Galactic bulge (GB)}
\label{sec:GB}

Whereas optical observations are prohibitive for the above two components,
the NSC and NSD/CMZ,
owing to high extinction ($A_V \gtrsim 30$~mag), the GB can be 
observed in the optical regime, at least at longer wavelengths.
This has allowed many large-scale surveys in the optical,
together with those in the infrared,
to produce massive datasets which have triggered rapid
growth in our knowledge, especially in the recent decades
(see, e.g.\  the contribution by Valenti in this volume).
In the low-latitude regions, however, optical observations
are very limited, and
strong extinction at $|b|\lesssim 1^\circ$ causes serious problems, even in the near infrared,
as we see in Section~\ref{sec:ExtLaw}.

The GB has an elongated structure,
although stars with different ages may be distributed
differently, and kinematics and chemical abundances of stars
show systematic trends depending on the position
(see, e.g.\  \cite[Ness \& Freeman, 2016]{Ness-2016a};
\cite[Portail \etal\ 2017]{Portail-2017};
\cite[Zoccali \etal\ 2017]{Zoccali-2017}).
Whereas the GB itself can be called a bar with its elongated structure,
other bars, longer and thinner than the GB, have been suggested,
although the situation is not so clear
(see reviews by \cite[Bland-Hawthorn \& Gerhard, 2016]{BlandHawthorn-2016},
and \cite[Zoccali \& Valenti, 2016]{Zoccali-2016}).
It is predicted that orbits along the bar structure produce
{\it cold} and high-$v_{\rm LoS}$ features
and such orbits may be preferentially
occupied by young bar stars, age $\lesssim$2--3~Gyr (\cite[Aumer \& Sch\"{o}nrich, 2015]{Aumer-2015}).
OH/IR stars and SiO maser sources
show kinematics expected for the bar-like orbits
(\cite[Habing, 2016]{Habing-2016}),
while high velocity components seen in
APOGEE's data may also be following
the expected bar kinematics (\cite[Palicio \etal\ 2018]{Palicio-2018}, and references therein). 
Among many new findings regarding the GB
based on various large-scale surveys (\cite[Barbuy \etal\ 2018]{Barbuy-2018}),
in the following discussions,
we focus on two important issues concerned with stellar populations in the GB:
namely, the presence of intermediate-age stars and the X-shaped structure,
both of which remain to be fully confirmed.

{\underline{\it Intermediate-age stars in the GB}}.
\\
Canonical results mainly based on photometric data suggest
that the bulge is composed almost purely of old stars, $\gtrsim 10$~Gyr
(\cite[Kuijken \& Rich, 2002]{Kuijken-2002}; \cite[Zoccali \etal\ 2003]{Zoccali-2003}; \cite[Clarckson \etal\ 2011]{Clarkson-2011}).
In contrast, Bensby \etal\ (\cite[2013]{Bensby-2013}, \cite[2017]{Bensby-2017})
obtained high-resolution spectra of microlensed dwarfs and subgiants
and, in the 2017 paper, they concluded that more than 35~\%
of metal-rich stars, [Fe/H]$>0$, are younger than 8~Gyr.
\cite{Bensby-2017} also found that 
such younger stars show {\it hot} kinematics like older GB stars.
The presence of the intermediate-age stars
was supported by \cite{Bernard-2018}, who used
an intensive dataset collected by HST/WFC3 and concluded that
20--25~\% of metal-rich stars (or 10~\% of the bulge stars)
are younger than 5~Gyr.
However, using similar or even larger datasets, \cite{Clarkson-2011}
and \cite{Renzini-2018} found no evidence of a significant
population of young ($\lesssim 5$~Gyr) stars.
It has been suggested that old stellar populations with
enhanced helium abundance may resolve such a conflict,
although an anomalously high helium enrichment seems to be required
(\cite[Nataf \& Gould, 2012]{Nataf-2012}; \cite[Nataf, 2016]{Nataf-2016}; \cite[Renzini \etal\ 2018]{Renzini-2018}).
Some types of bright evolved stars may contribute to 
this issue if one can give accurate ages for those
in the last evolutionary phases with short time scales.
For example, mass-losing AGB stars with maser emission
(\cite[Blommaert \etal\ 2018]{Blommaert-2018}; \cite[Trapp \etal\ 2018]{Trapp-2018}) and
carbon-rich stars 
(\cite[Matsunaga \etal\ 2017]{Matsunaga-2017}) often
indicate intermediate-age populations. However,
the methods of determining the ages of such objects 
have not been fully established. Moreover,
short evolutionary phases are expected to give a relatively
small number of objects, and the interpretation of such rare objects
requires caution considering the presence of blue star stragglers
(\cite[Clarkson \etal\ 2011]{Clarkson-2011}),
which are old but may mimic objects evolved from heavier single stars.

{\underline{\it Double red clump and the X shape of the GB}}.
\\
Using large-scale 2MASS and/or OGLE data, \cite{McWilliam-2010} and
\cite{Nataf-2010}
detected two split peaks of the red clump (or a double RC)
at high Galactic latitudes, $|b|\gtrsim 5^\circ$, in the GB.
Many studies have concluded that this feature reflects
the X shape that characterises a peanut-shaped bulge 
(\cite[Zoccali \& Valenti, 2016]{Zoccali-2016}, and references therein).
Some works claim that the double RC can be explained
by the presence of helium-enhanced populations
(Lee \etal\ \cite[2015]{Lee-2015}, \cite[2018]{Lee-2018}).
There are increasing publications concerning
the double RC and the X shape of the GB (e.g.\ 
\cite[Ness \etal\ 2012]{Ness-2012};
\cite[Wegg \& Gerhard, 2013]{Wegg-2013};
\cite[Gonzalez \etal\ 2015]{Gonzalez-2015};
\cite[Ness \& Lang, 2016]{Ness-2016b};
L\'{o}pez-Corredoira, \cite[2016]{LopezCorredoira-2016}, \cite[2017]{LopezCorredoira-2017};
\cite[Joo \etal\ 2017]{Joo-2017}).
See also the point raised by E.~Valenti, which is given
in the {\it Discussion} section of this contribution.
It is worthwhile to note at least that
different stellar populations have different distributions.
For example, the old (and probably metal-poor) population
represented by RR Lyrae, type II Cepheids, and short-period Miras do not show
evidence of the X-shaped structure
(\cite[Pietrukowicz \etal\ 2013]{Pietrukowicz-2013}; \cite[Catchpole \etal\ 2016]{Catchpole-2016}; \cite[Bhardwaj \etal\ 2017]{Bhardwaj-2017}; \cite[Braga \etal\ 2018]{Braga-2018}).
It is therefore important to understand the parameters
(age, metallicity, and helium abundance if possible;
see Section~\ref{sec:He}) of stars being investigated.

\subsection{Extinction towards the GC}
\label{sec:ExtLaw}

It is well known that regions towards the GC are severely obscured
by interstellar extinction, and observations of stars therein
are limited or complicated by its effects. 
For example, the Optical Gravitational Lensing Experiment (OGLE)
has discovered over 38,000 RR Lyrae in a large area towards the GB,
but very few were found within {$\sim$}1.5~degrees of the Galactic plane
(\cite[Soszynski \etal\ 2014]{Soszynski-2014}).
More recently, {\it Gaia} Data Release 2 (DR2) compiled
the first catalogue from the all-sky variability search
(\cite[Gaia Collaboration, 2018b]{Gaia-2018b}; \cite[Holl \etal\ 2018]{Holl-2018}),
and \cite{Mowlavi-2018}, studying long-period variables,
found that the low-latitude
region ($|b|\lesssim 1^\circ$) is the zone of avoidance (see their Fig.~46).
According to the extinction map of \cite{Dutra-2003}, the extinction
at {$\sim$}2~{$\mu$}m, $A_K$\footnote{Generally speaking,
it is important to avoid confusion between the $K$ and $K_{\rm s}$ bands (see \cite[Bessell, 2005]{Bessell-2005}, for various photometric bands). The difference, however, has no significant effect on the discussions in this contribution, and we replace $K_{\rm s}$ by $K$ anywhere relevant.}, gets higher than 1~mag
in such a low-latitude region.
Infrared observations are much less affected,
but there are still challenging issues.
Strong extinction tends to render
stars such as RR Lyrae, which are not so luminous intrinsically,
below detection limits of surveys 
(\cite[Contreras Ramos \etal\ 2018]{ContrerasRamos-2018}).
Once reddened objects are detected, it is crucial to accurately know
the extinction law, i.e.\  the wavelength dependency of the extinction,
to discuss the distribution of stellar populations
in highly obscured regions such as the inner Galaxy
(see the reviews of \cite[Matsunaga, 2017]{Matsunaga-2017b}, and  \cite[Matsunaga \etal\  2018]{Matsunaga-2018}).

It is important to keep in mind that the extinction towards
the GC is very patchy.
For example, \cite{Matsunaga-2009} estimated the reddenings
and extinctions of short-period ($<350$~days) Miras
within ${\sim}0.3^\circ$ around the GC. 
In their study, although {$\sim$}150 Miras with distance and $A_K$ obtained
are projected towards the NSD/CMZ and most of them are
located at almost the same distance, 8.24~kpc,
the precision of individual distances is not high enough
to determine whether the Miras belong to the NSD/CMZ or to the GB.
The derived extinctions
range from {$\sim$}2~mag to more than 3.5~mag and 
vary with the celestial position
within a small angular distance; 
in some cases, two Miras separated by only a couple of arc-minutes
have $A_K$ different by more than 1~mag
(see Fig.~25 in \cite[Matsunaga \etal\ 2009]{Matsunaga-2009}).
Whereas foreground molecular clouds clearly cause
the patchy extinction pattern, resulting in some regions
being free of any Miras found at the bulge distance, 
part of the large scatter in $A_K$ may be attributed to 
the dust within the NSD/CMZ. It is possible that stars in its front are
less reddened compared to those on the far side, although 
distances to individual objects are not precise enough to
confirm such a systematic effect. There are a few three-dimensional
extinction maps which give the extinction as a function of
the celestial coordinate and distance, and that by
\cite{Schultheis-2014}, who used VVV and 2MASS data,
is particularly useful for the regions
around the GC. The authors found that a large amount of
dust exists in front of the Galactic bar, at 5--7~kpc,
but the resolution was insufficient to discuss
the extinction within or near the NSD/CMZ
(or the data were not deep enough).
Nevertheless, a couple of studies have suggested
a significant amount of extinction
within the NSD (\cite[Launhardt \etal\ 2002]{Launhardt-2002};
\cite[Oka \etal\ 2005]{Oka-2005}).
For the NSC, \cite{Chatzopoulos-2015b} performed a detailed study
on the kinematics of stars that seems to be skewed owing to
the selection bias caused by the additional extinction,
and concluded that the dust within the NSC is responsible for 
{$\sim$}15~\%, $A_K\simeq 0.4$~mag, of the total extinction.
They demonstrated that the distribution and kinematics of
stars can provide information on the distribution of
the absorbing ISM (see also Section~\ref{sec:Astrometry}).

\section{Approaches for disentangling mixed populations in the GC}
\label{sec:Approach}

In order to address the issues we discussed in Section~\ref{sec:Review},
there are certain observational challenges. 
An obvious advantage of ELTs is the deeper limiting magnitudes, but
it will be important to
combine the deep ELT data with various kinds of other observational data
when we confront the challenges, as discussed below.
First, when observed objects are located in the LoS with
many overlapping components, e.g.\  the NSD, GB, and foreground/background disc,
it is necessary to know which system
individual objects belong to. Furthermore, it would be
useful to locate the objects inside the system in question.
It has been, however, very hard to determine locations
within the NSD with the radius of {$\sim$}250~pc
(0.07~mag in distance modulus), e.g.\  by taking the correction of
the severe extinction into account.
Astrometric data leading to three-dimensional velocities
when combined with radial velocities would allow us to
locate stars, at least roughly, within the NSD
(Section~\ref{sec:Astrometry}).
In addition, the detection of absorption features caused by
the ISM would provide valuable information on
the locations of both stars and the intervening ISM
(Section~\ref{sec:Interstellar}). 
This latter approach can also help us deal with the second challenge:
we need to estimate the foreground extinction for each object
because of the very patchy nature of the extinction towards the GC (Section~\ref{sec:Interstellar}).
The third challenge lies in carrying out spectroscopic 
measurements of faint dwarfs and subgiants in the GB as much as possible,
for which the ELTs' high sensitivity is of direct importance,
and we will also discuss the merits of microlensing surveys in the infrared
in Section~\ref{sec:Dwarfs}.
Finally, for both topics raised for the GB
(Section~\ref{sec:GB}), helium abundance
may play a key role in understanding the important characteristics
of the relevant stellar populations.
It would be of great value if we could measure the helium abundances
of stars in the GB, and we will discuss the possibility of
direct measurements using a helium line in the near infrared
in Section~\ref{sec:He}, although it is not entirely clear how easily
and accurately the abundances can be measured.

\subsection{Astrometric data}
\label{sec:Astrometry}

A common problem in understanding objects in the NSD is,
as mentioned above,
that the precision of distances to them is usually not
high enough to determine where along the LoS,
e.g.\  on which of the near and far sides, they are located,
even if one could identify objects belonging to the NSD.
It is usually impossible to reduce the distance errors 
to $\pm 250$~pc at the distance of {$\sim$}8.3~kpc
(i.e.\  3~\%), taking into account the uncertainty in
the foreground extinction and other errors.
Additionally, such a precision (3.5~{$\mu$}as in parallax)
is very difficult to achieve, which will likely be the case  even for future astrometric
observations (either infrared or radio wavelengths).
Radial velocities, $v_{\rm LoS}$,
leave an ambiguity between near and far distances,
even if we consider circular orbits only.
To overcome this problem, proper motions are expected to be very useful.
Precise proper motions, especially those in the $b$ direction,
allow us to clean a sample of stars in the NSD/CMZ ($\mu _{b}\sim 0$),
which is otherwise severely contaminated by stars in the GB 
with a large velocity dispersion,
$\sigma (\mu _{b}) \gtrsim 3$~mas\,yr$^{-1}$ (\cite[Soto \etal\ 2014]{Soto-2014}).
A precision of {$\sim$}0.3~mas\,yr$^{-1}$ is desirable for selecting
the NSD/CMZ stars efficiently. 
In contrast, proper motions in the $l$ direction, $\mu_{l*}$, would
enable us to locate stars inside the NSD/CMZ, because 
stars on the near side and those on the far side
are expected to have opposite tangential velocities owing to
the rotation within the NSD/CMZ. 
For example, \cite{Chatzopoulos-2015b} made use of $\mu_{l*}$ as a proxy of
the far--near positions of stars inside the NSC, 
following a similar idea but based on a sophisticated model of the cluster,
and estimated the extinction caused by dust within the NSC.
Similar approaches can be used for the NSD/CMZ stars when
the necessary astrometric data become available.
Considerable uncertainties
can remain in the locations of the stars,
considering the possibility of {\it open} orbits,
but it is possible to distinguish the near and far distances
as long as we assume that the objects are not counter-rotating in
the NSD/CMZ (\cite[Stolte \etal\ 2014]{Stolte-2014}; \cite[Kruijssen \etal\ 2015]{Kruijssen-2015}).
The use of $v_{\rm LoS}$ and $\mu_{l*}$
is also discussed in detail by \cite{Debattista-2018}.
In particular, their Figure~3 shows predicted trends of the velocities
as functions of the position around the GC for
two groups with different ages and different kinematics
(one dominated by stars streaming along the bar and
the other for those in the nuclear disc).

Long-term monitoring surveys from the ground allow us to measure
the proper motions of stars even though the precision is not so high as
that of astrometric satellites. 
As an early achievement, \cite{Sumi-2003} detected the streaming motion of
the Galactic bar by measuring the proper motions of red clump giants
based on four-season, 1997--2000, data from OGLE-II. The precision of
the proper motion was typically 1--2~mas\,yr$^{-1}$ for individual stars
(\cite[Sumi \etal\ 2004]{Sumi-2004}),
and using thousands of stars allowed the detection of
the tangential streaming motion of {$\sim$}1.6~mas\,yr$^{-1}$, corresponding
to {$\sim$}100~km\,s$^{-1}$ (see also follow-up studies by
\cite[McWilliam \& Zoccali, 2010]{McWilliam-2010}, and \cite[Poleski \etal\ 2013]{Poleski-2013}).
Recently, \cite{Smith-2018} compiled a catalogue of
astrometric measurements based on VVV survey data,
and they obtained proper motions with statistical uncertainties
below 1~mas\,yr$^{-1}$ for nearly 50 million stars.
These measurements are useful to study the kinematics of stars
around the GC, although the precision is not very high for
individual stars. 
Similar or higher precision can be achieved with AO-applied images
(e.g.\  \cite[Stolte \etal\ 2014]{Stolte-2014}) with a smaller number of epochs,
but such observations cover smaller fields-of-view than
the large-scale surveys mentioned above.
Compared to ground-based measurements, 
space projects with astrometric satellites have a clear advantage
in terms of the precision.
{\it Gaia} has been providing very precise measurements of proper motions,
in addition to parallaxes, with the median error of 50~{$\mu$}as\,yr$^{-1}$ in the DR2
for a large number of stars in the Galaxy (\cite[Gaia Collaboration, 2018a]{Gaia-2018a}; \cite[Lindegren \etal\ 2018]{Lindegren-2018}),
and the errors will be smaller than 10~{$\mu$}as\,yr$^{-1}$ for sufficiently bright stars at the end of mission. However, 
its observation
in the optical wavelengths cannot reach the inner regions of the Galaxy.
There are a few plans which can give space-borne astrometric
measurements in the infrared, namely
Small-JASMINE\footnote{The broad band, $H_{\rm W}$, designed for the Small-JASMINE cover 1.1--1.7~{$\mu$}m, and its bulge survey area covers
approximately 2 deg$^2$ towards the NSD/CMZ.}
(\cite[Gouda, 2018]{Gouda-2018}),
WFIRST (\cite[WFIRST Astrometry Working Group, 2017]{WFIRST-2017}),
and GaiaNIR (\cite[Hobbs \etal\ 2016]{Hobbs-2016}).
At the distance of 8~kpc, the precisions of {$\sim$}0.5~mas\,yr$^{-1}$ (ground-based) and $\lesssim 25$~{$\mu$}as\,yr$^{-1}$ (Small-JASMINE)
correspond to the tangential velocities of 20 and 1~km\,s$^{-1}$, respectively.
As for Small-JASMINE,
in addition to bright main objects
(approximately 7,000 bulge stars with $H_{\rm W}<12.5$) for which the precision of
25~{$\mu$}as\,yr$^{-1}$ or higher is targeted,
there is a plan to observe fainter 
objects ({$\sim$}70,000 bulge stars with $H_{\rm W}<15$)
with lower precision {$\sim$}125~{$\mu$}as\,yr$^{-1}$.
This intermediate precision,
which corresponds to 5~km\,s$^{-1}$ at the GC distance,
will be useful for filling the gap between the objects with the best
proper motions and those with ground-based astrometry.

\subsection{Interstellar absorption features}
\label{sec:Interstellar}

Interstellar clouds not only attenuate signals
to be accounted for (broad-)band magnitudes, but also imprint
absorption features in stellar spectra.
Such absorptions reflect the amount of extinction
and the velocity structure of the absorbing materials.
Here, we here discuss diffuse interstellar bands (DIBs) and
absorption lines by certain molecules that have been observed 
in stars, particularly those in the direction of the GC.

DIBs can be useful to determine foreground extinctions of
individual objects.
Some DIBs in the optical wavelengths have been found to
show good correlation between their equivalent widths (EWs) and the extinction
(e.g.\  \cite[Kashuba \etal\ 2016]{Kashuba-2016}).
In contrast, at least some other lines show
the so-called skin effect which gives scatters around simple reddening--EW(DIB) relations
depending on the atomic/molecular ratio along intervening clouds (\cite[Herbig, 1995]{Herbig-1995}; \cite[Lan \etal\ 2015]{Lan-2015}). 
In any case, the ranges of the extinction
investigated by using optical DIBs
are limited, compared to
the extinction towards the GC regions,
in previous studies.
\cite{Damineli-2017}
extended the relationship between the {8,620\,{\AA} DIB and the extinction
up to $A_K \sim 0.8$~mag by observing stars in
the famous massive cluster Westerlund~1 in the inner Galaxy
($l=-20.5^\circ, b=-0.4^\circ$). 
This particular DIB feature will be especially important, because
it is included within the wavelength range of {\it Gaia}'s
radial velocity spectrometer (RVS). For severely reddened stars
in the innermost part of the Galaxy, it is necessary to identify
DIBs in the infrared range showing good correlations with the extinction.
For example, \cite{Cox-2014} detected more than 20 DIBs
in the wide range of the NIR, 0.9--2.5\,{$\mu$}m, observed
with the VLT/X-Shooter,
including 11 candidate DIBs they newly identified.
Later,
Hamano \etal\ (\cite[2015]{Hamano-2015}, \cite[2016]{Hamano-2016}) found nearly 20 DIBs at 0.98--1.32~{$\mu$}m
with WINERED spectra ($R\sim 28,000$),
and \cite{Galazutdinov-2017} detected 14 DIBs
at 1.46--2.40~{$\mu$}m with IGRINS spectra ($R\sim 45,000$).
Both spectrographs used for these works can produce
high-quality spectra (S/N$\gtrsim 500$),
which is important for the detection of many shallow DIBs,
down to {{$\sim$}5~m{\AA}} with WINERED
or {{$\sim$}30~m{\AA}} with IGRINS.
We can expect that DIBs in severely reddened stars are relatively stronger,
and in fact \cite{Geballe-2011} detected 13 DIBs
in a few objects located within the NSD/CMZ
at 1.5--1.8~{$\mu$}m in the $H$ band only.
Nevertheless, it is important to obtain high-resolution and
high-quality spectra especially if one would like to know
the velocity profile of the DIB absorption.
A comprehensive review of DIBs is found in
\cite{Krelowski-2018}, which mainly discusses the optical DIBs but
also includes the infrared ones.

The relations between the DIBs' EWs and the extinction, once established, 
can be used to estimate the extinctions of individual objects.
Such estimates would be independent of the distances of objects
and work as an alternative or complementary method
to photometry-based methods.
Furthermore, DIBs can provide valuable information on
the velocity distribution of absorbing ISM.
For example,
DIBs produced by the ISM in the NSD/CMZ are characterised by
a large velocity dispersion, {$\sim$}150~km\,s$^{-1}$
(\cite[Geballe \etal\ 2011]{Geballe-2011}),
in contrast to most of the foreground ISM
which has almost zero velocity in the direction of the GC.
Recently, making use of the large dataset of the APOGEE,
\cite{Zasowski-2015} found that the DIB with  
$\lambda_0=${15,273\,\AA}, according to their determination, 
shows red- (or blue-) shifts following the Galactic rotation
(see also
Elyajouri \etal\ \cite[2016]{Elyajouri-2016}, \cite[2017]{Elyajouri-2017},
for DIBs in APOGEE spectra).
It should be noted that the FWHMs of DIBs tend to be large
compared to those of lines of interstellar atoms and molecules,
which we discuss below.
In addition, the profile of the band and even the central wavelength
may be subject to the physical conditions of the absorbing ISM
(\cite[Cami \etal\ 2004]{Cami-2004}; \cite[Oka \etal\ 2013]{Oka-2013}).
These can blur and/or bias the velocity distribution inferred from DIBs.

Different types of absorption features caused by ISM
are also useful.
For example, the absorption lines of ${\rm H}_3^{+}$
in the $L$ band have been intensively investigated with the spectra of
early-type stars located around the GC
(\cite[Goto \etal\ 2002]{Goto-2002}; \cite[Oka \etal\ 2005]{Oka-2005},
and references therein).
The advantage of these lines is their sharpness, which makes it possible
to identify velocity components with high resolution as demonstrated
in the above works.
Furthermore, 
the ${\rm H}_3^{+}$ absorptions in different transitions
enable us to investigate the physical and chemical conditions of
the ISM in the NSD/CMZ and elsewhere
(\cite[Goto \etal\ 2014]{Goto-2014}; \cite[Le Petit \etal\ 2016]{LePetit-2016}).
Other useful features for studying interstellar absorption include
the lines of CO and the less-investigated H$_2$
(e.g.\  Goto \etal\ \cite[2014]{Goto-2014}, \cite[2015]{Goto-2015}; \cite[Lacy \etal\ 2017]{Lacy-2017})
in the $K$ band, and
combining these features would allow further detailed investigations
of the chemical status of the ISM.

The interstellar features discussed here are
rather weak in most cases, and
it is crucial to collect high-quality spectra
(i.e.\  high resolution and high S/N).
Whereas some features are detectable with the resolution of {$\sim$}20,000,
e.g.\  the DIB studied with the APOGEE, higher resolving power
($R\gtrsim 50,000$) is desirable considering the sharpness of
the interstellar absorption lines. 
Their strengths do not exceed 10~\% in depth, even in the spectra
of stars at around the GC, thus requiring high S/N of a few hundred.
DIBs in the spectra of cool stars 
(\cite[Monreal-Ibero \& Lallement, 2017]{MonrealIbero-2017})
may be very useful for studying the ISM towards the GC regions, but there are
a few issues: large errors in the databases of stellar lines in the IR,
and diversity of the objects in temperature and chemical abundances. 
Relative spectral analyses comparing stars in the NSD/CMZ and the inner GB
with those which have similar parameters but significantly
weaker DIBs in the outer GB may make it easier to measure
the DIB components. After all, the GC regions are full of cool stars,
and it will be possible to find very similar stars.
Studying DIBs in Cepheids is also interesting, especially those
located in the disc, considering that 
individual Cepheids give good estimates of distance and reddening
thanks to the period--luminosity and period--colour relations
(\cite[Kashuba \etal\ 2016]{Kashuba-2016}),
although the task may be even more challenging owing to the pulsation
causing spectral variation of each Cepheid (e.g.\  offsets and
asymmetric profiles of lines). 

\subsection{Spectroscopy for bulge dwarfs/subgiants with and without microlensing events}
\label{sec:Dwarfs}

As described in Section~\ref{sec:GB}, an important question about the age distribution
of stellar populations in the GB came with high-resolution
spectroscopic observations of microlensed dwarfs and subgiants
(\cite[Bensby \etal\ 2013]{Bensby-2013}). They are too faint for current
high-resolution spectrographs with 8-m class telescopes,
but the microlensing events with amplitudes of 10--2,000
sufficiently brightened the targets. The high sensitivity of ELTs will
enable us to observe dwarfs and subgiants in the GB
even without the microlensing events as we evaluate in Section~\ref{sec:ELTs}.

Most of the previously detected microlensing events are
limited by the sensitivity of optical microlensing surveys,
and they are away from the Galactic plane by 1.5 degrees or more.
In contrast, a pioneering survey using UKIRT successfully detected
microlensing events at low Galactic latitudes,
$-0.98 \leq b \leq -0.36$~degrees (\cite[Shvartzvald \etal\ 2017]{Shvartzvald-2017}).
\cite{Navarro-2017} further reported nearly 200 microlensing events
detected in the VVV survey for five years in {$\sim$}5~deg$^2$ around the GC.
A great jump in infrared microlensing surveys will be delivered
by the WFIRST project being planned for the mid-2020s
(\cite[Bennett \etal\ 2018]{Bennett-2018}).
There are also plans for ground-based
microlensing surveys in the infrared including
PRIME (a new 1.3-m telescope with a 1.3-deg$^2$ wide-field camera being constructed
in South Africa) in addition to the currently active infrared facilities,
UKIRT and VISTA
(\cite[Yee \etal\ 2018]{Yee-2018}).
Such surveys in the future will
allow us to study dwarfs and subgiants in the low-latitude regions
by using the same technique as \cite{Bensby-2013} and other studies,
but based on infrared photometric and spectroscopic observations.

\subsection{He\,I {10,830~{\AA}} and helium abundance}
\label{sec:He}

Despite its importance in stellar structure and evolution, measurements 
of helium abundance using helium line(s)
have not been performed particularly intensively.
Direct measurements of helium abundance are
not so easy, as we discuss below. Readers are 
also referred to valuable reviews in \cite{Gratton-2012}
and \cite{Bastian-2018}. There are some methods
for indirectly measuring the helium abundances of stellar populations
based on photometric data,
e.g.\  the so-called $R$ parameter
(\cite[Iben, 1968]{Iben-1968}; \cite[Salaris \etal\ 2004]{Salaris-2004})
and the properties of RR Lyrae
(\cite[Marconi \& Minniti, 2018]{Marconi-2018a}; \cite[Marconi \etal\ 2018]{Marconi-2018b}),
but the direct measurements
are of great value for investigating the mixed populations
in the GC regions.

In shorter optical wavelengths ({$< 5,900$~\AA}),
some photospheric He\,I lines can be detected, but
they appear only in hot stars ($\gtrsim 9,000$~K). 
Such hot stars in old stellar populations appear
during the horizontal branch (HB) phase,
but radiative levitation or gravitational settling of various elements
can occur in the turbulence-free atmosphere of sufficiently hot HB stars
(\cite[Grundahl \etal\ 1999]{Grundahl-1999}; \cite[Behr \etal\ 1999]{Behr-1999}; \cite[Behr, 2003]{Behr-2003}).
This process produces the $u$ jump observed in colour--magnitude diagrams:
the hotter stars appear distinctly bright in the Str\"{o}mgren $u$ band 
(\cite[Grundahl \etal\ 1999]{Grundahl-1999}). 
Fortunately, for HB stars in the narrow temperature range of
$9,000 \lesssim T_{\rm eff} \lesssim 11,500$~K,
the He abundance can be measured
without the diffusion depletion, and such measurements have been reported for
several globular clusters using the photospheric He\,I line at {5875~{\AA}}
(\cite[Gratton \etal\ 2014]{Gratton-2014};
\cite[Mucciarelli \etal\ 2014]{Mucciarelli-2014},
and references therein). The temperature of the $u$ jump
(or the {\it Grundahl} jump) has been found to be
{$\sim$} 11,500~K for most globular clusters, but it can also vary with
the helium abundance as suggested by \cite{Tailo-2017}.
Unfortunately, these photospheric He\,I lines in the optical regime
are of limited use for stars in the GC regions.
First, the hot HB stars are faint at ${\sim}$8~kpc for
high-resolution spectroscopy, and
the interstellar extinction makes necessary optical observations
prohibitive, except for low-extinction regions of the outer bulge.
Second, with the high metallicity of the bulge stars,
HB is no longer extended to the hot temperature range.

The He\,I {10,830~{\AA}} line\footnote{The He\,I transition at {10,830~{\AA}} (in the air wavelength scale) is a triplet composed of two unresolved components at {10,830.30~{\AA}} and an insignificant component at {10829.08~{\AA}} (\cite[Dupree \etal\ 1992]{Dupree-1992}). In the vacuum scale, the wavelength corresponds to {$\sim$}10,833~{\AA}.}
exists 
in the near-infrared range,
and this line is observable in FGK giants (Dupree \etal\ \cite[1992]{Dupree-1992}, \cite[2009]{Dupree-2009}). 
It has been used in several studies
on the helium abundances of multiple populations in globular clusters
(\cite[Dupree \etal\ 2011]{Dupree-2011}; \cite[Pasquini \etal\ 2011]{Pasquini-2011}; \cite[Dupree \& Avrett, 2013]{Dupree-2013}; \cite[Smith \etal\ 2014]{Smith-2014}; \cite[Strader \etal\ 2015]{Strader-2015}).
An important point which should be kept in mind about this line is,
however, that it is formed in stellar chromospheres, and 
the abundances cannot be obtained without careful, challenging modelling
of the chromosphere (\cite[Strader \etal\ 2015]{Strader-2015}).
In addition to many other scientific targets
(e.g.\  \cite[Izotov \etal\ 2014]{Izotov-2014}; \cite[Aver \etal\ 2015]{Aver-2015}; \cite[Spake \etal\ 2018]{Spake-2018}),
this line can be
used for a variety of applications, such as the dynamics in
the envelopes of different types of objects,
e.g.\  the Sun (\cite[Dupree \etal\ 2005]{Dupree-2005}), metal-poor giants (\cite[Dupree \etal\ 2009]{Dupree-2009}),
metal-poor dwarfs (\cite[Takeda \& Takada-Hidai, 2011]{Takeda-2011}),
active stars (\cite[Sanz-Forcada \& Dupree, 2008]{SanzForcada-2008}; \cite[Fuhrmeister \etal\ 2018]{Fuhrmeister-2018}), and young stellar objects (\cite[Takami \etal\ 2002]{Takami-2002}; \cite[Connelley \& Greene, 2010]{Connelley-2010}; \cite[Reipurth \etal\ 2010]{Reipurth-2010}; \cite[Cauley \& Johns-Krull, 2014]{Cauley-2014}). 
Combined, these applications provide strong motivation
to establish the usage of the He\,I {10,830~{\AA}} in general and
the chromosphere models in particular for predicting
the formation of this line.

It should be noted that, apart from the theoretical challenge
of modelling the chromosphere,
there are certain observational challenges in investigating
the He\,I {10,830~{\AA}} line of red giants in the GC regions.
First, there are a few telluric absorption lines
around this wavelength (see, e.g.\  \cite[Sameshima \etal\ 2018]{Sameshima-2018}),
and they should be accurately removed. 
Second, ThAr lamps, which are most often used for wavelength calibration, do not show many lines
around the wavelength (\cite[Kerber \etal\ 2008]{Kerber-2008});
a low-accuracy dispersion solution can cause a problem,
especially if the He line is weak and blended with a telluric line.
A practical approach is to use late-type giants with established radial velocity as wavelength standards
(e.g.\  \cite[Dupree \etal\ 2011]{Dupree-2011}),
but new calibration sources under development,
e.g.\  lamps with other elements such as uranium (\cite[Redman \etal\ 2011]{Redman-2011}) and
laser frequency combs (\cite[Osterman \etal\ 2012]{Osterman-2012}),
will eventually enable accurate and robust calibration.
Third, the He\,I line is expected to be weak or
invisible in the main bulk of bright red giants in the bulge.
It gets strongest, $\gtrsim 300$~m{\AA},
at the spectral types of G and early K,
and becomes weaker, {$\lesssim$100~m{\AA} at 5,000~K or lower,
as effective temperature decreases
(\cite[Vaughan \& Zirin, 1968]{Vaughan-1968}; \cite[Strader \etal\ 2015]{Strader-2015}).
Therefore, we need to collect high-resolution spectra of
early-K giants (or those with even earlier spectral type)
or late-K giants with extremely high S/N. 
The effective temperature of red clump giants is approximately 5,000~K,
which indicates that the main targets have $J < 15$
even in the lowest-extinction regions of the GB.
This is roughly the limit of the current good spectrographs
attached to 8-m class telescopes, considering the required high quality,
but ELTs will reach such targets,
as we discuss in Section~\ref{sec:ELTs}.

\section{Observational opportunities with ELTs}
\label{sec:ELTs}

In this section, we briefly discuss what types of tracers 
in different regions around the GC can be reached with
the expected limiting magnitudes of ELTs. 
We focus on high-resolution near-infrared spectroscopy
($R\equiv \lambda/\Delta\lambda \gtrsim 20,000$).
According to \cite{Zieleniewski-2015}, for example,
HARMONI attached to E-ELT is expected to reach
approximately $J=19.5$ and $K=19.5$, in the Vega magnitude scale,
for $R=20,000$
with S/N$=$50 for five hours on-source.
We use these limits as a guideline in the following discussion.
It is crucial to collect high-quality (high-resolution and high-S/N) spectra for many science targets discussed in this contribution.
In particular, studying DIBs
and lines of interstellar molecules requires S/N as high as a few hundred, 
as mentioned in Section~\ref{sec:Interstellar},
and the limiting magnitudes for the relevant projects would 
probably be {$\sim$}2~mag shallower.
Such a high requirement means
that the high photon-collecting powers of ELTs will be higly
advantageous once suitable high-resolution spectrographs are installed.
High-quality spectroscopy will be also important
for measuring the abundances of extremely metal-poor stars
in the bulge
(see, e.g.\  Howes \etal\, \cite[2015]{Howes-2015}, \cite[2016]{Howes-2016}),
once such candidates are identified, although the necessary quality
for such projects is less demanding than the interstellar absorption features.

Figure~\ref{fig1} plots isochrones offset by 
the distance modulus of the GC, $14.6$~mag, as well as
by different reddenings and extinctions
($A_K=0.5, 1.5$, and 2.5~mag),
following the reddening vector of \cite{Nishiyama-2006}.
The isochrones for 10~Gyr (solid curve) and 100~Myr (dashed curve),
both with solar metallicity, are taken from 
the MESA Isochrones \& Stellar Tracks (MIST) website
(\cite[Choi \etal\ 2016]{Choi-2016}),
and the evolutionary phases
prior to thermal-pulsing AGB are included.
Upper main-sequence stars of young stellar populations
are good targets for highly precise detection of interstellar
absorption features in the $K$ band and longer wavelengths,
even in the heavily obscured regions, $A_K \gtrsim 3$~mag (Section~\ref{sec:Interstellar}).
The typical positions of the red clump at different
extinctions ($A_K=0, 1, 2$, and 3~mag) 
are also indicated by squares connected by the line
whose slope corresponds to that of the reddening vector.
Comparing the two panels clearly shows that
objects quickly get faint in $J$ more than in $K$ with increasing extinction. 
For example, the RC in the NSD/CMZ is not available for
high-resolution spectroscopy in $J$ but observable in $K$.

\begin{figure}[b]
\begin{center}
\includegraphics[width=0.98\hsize]{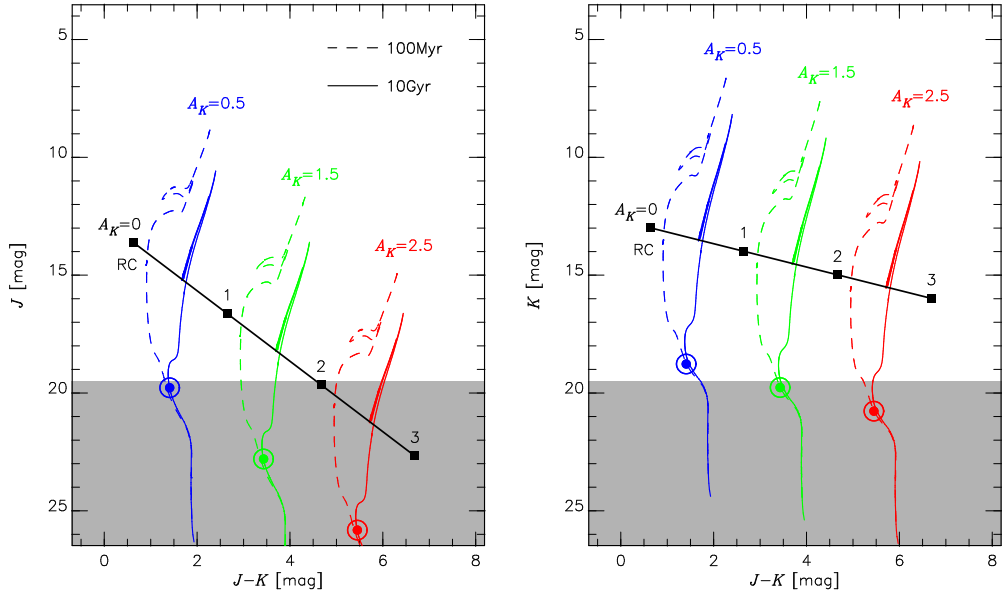} 
 \caption{Isochrones shifted down to the distance of
 the GC ($\mu_0=14.6$~mag) but with different reddenings and extinctions, $A_K$,
 added as indicated in the colour--magnitude diagram:
 $(J-K)$--$J$ (left) and $(J-K)$--$K$ (right).
 The selected extinctions, $A_K=0.5, 1.5$, and 2.5~mag,
 correspond to the typical values for the high-latitude GB ($|b|\gtrsim 1^\circ$), the low-latitude GB, and the NSD/CMZ, respectively.
 The positions of the red clump (square; \cite[Laney \etal\ 2012]{Laney-2012}) and the Sun ($\odot$ symbol; \cite[Willmer \etal\ 2018]{Willmer-2018}) with different $A_K$ values
 are also plotted. }
   \label{fig1}
\end{center}
\end{figure}

Table~\ref{tab1} lists the expected $VJK$ magnitudes 
of some types of objects at the distance of the GC. Considering the ranges
of the extinction, 
the expected magnitudes ranges are given for each combination of the object type and the photometric band for
three regions: high-latitude GB ($0.3\lesssim A_K \lesssim 1$),
low-latitude GB ($1 \lesssim A_K \lesssim 2$), and
the NSD/CMZ ($2 \lesssim A_K \lesssim 3.5$).
For comparison, the expected magnitude ranges in the optical $V$ band
are included. Some bright objects may be visible in the optical wavelengths
in low-extinction regions, but in high-extinction regions of the GB
and the NSD/CMZ, any tracers we discuss here become
invisible ({$\sim$}30~mag or fainter), even with ELTs.

\begin{table}
  \begin{center}
  \caption{Expected $VJK$ magnitudes of some types of objects in the GB and the NSD/CMZ.}
  \label{tab1}
 {\scriptsize
  \begin{tabular}{c|c|c|c|c}
  \hline 
  & Absolute mag. & GB ($|b|\gtrsim 1^\circ$) & GB ($|b|\gtrsim 1^\circ$) & NSD/CMZ \\
    $A_K$ &  & 0.3--1.0 & 1.0--2.0 & 2.0--3.5  \\
  \hline
    Miras    &  $-6>K>-8.5$   &  9.5--6.3 & 10.5--7.0  & 12--8  \\
             &  $-5>J>-6.5$           & 12.5--9.4 & 15.5--11.5 & 20--14.5  \\
             &  $+3>V>-2$             & $>17.3$ & $>28.5$ & $>45$  \\
  \hline
   RC giants &  $-1>K>-1.5$   & 14.5--13.3 & 15.5--14 & 17--15  \\
             &  $-0.5>J>-1$           & 17--14.4 & 20--16.5 & 24.5--19.5  \\
             &  $+1.5>V>+0.5$         & $>19.8$ & $>31$ & $>47$  \\
  \hline
    RR Lyrae &  $0>K>-1$      & 15.5--13.8 & 16.5--14.5 & 18--15.5  \\
             &  $0>J>-1$              & 17.5--14.4 & 24.7--20.9 & 25--19.5  \\
             &  $1>V>0$               & $>19.3$ & $>30.5$ & $>46.5$  \\
  \hline
    G dwarfs &  $+3.7>K>+3.0$ & 19.2--17.8 & 20.2--18.5 & 21.7--19.5  \\
             &  $+4.2>J>+3.4$         & 21.7--18.8 & 24.7--20.9 & $>23.9$  \\
             &  $+5.5>V>+4.5$         & $>23.8$    & $>35$ & $>51$  \\
  \hline 
  \end{tabular}
  }
 \end{center}
\vspace{1mm}
 \scriptsize{
 {\it Notes:}\\
  $^1$Absolute magnitudes of each tracers are taken from \cite{Ita-2004}, \cite{Marconi-2015}, \cite{Pecaut-2013}, \cite{Salaris-2002}, and \cite{Soszynski-2009}. \\
  $^2$The range of $A_K$ for each region is converted to those of $A_J$ and $A_V$, considering the following extinction ratios: $A_J/A_K=3.0$ (\cite[Nishiyama \etal\ 2006]{Nishiyama-2006}) and $A_V/A_K=16$ (\cite[Nishiyama \etal\ 2008]{Nishiyama-2008}).}
\end{table}

Figure~\ref{fig1} and Table~\ref{tab1} indicate that 
it is possible to take high-resolution spectra 
of dwarfs and subgiants even without microlensing events;
low-extinction, $A_K \lesssim 0.5$, regions of the GB
can be reached in $J$, whereas regions with higher extinction, $A_K \gtrsim 1$,
can be reached only in $K$ and longer wavelengths.
Therefore, we can observe a large number of dwarfs and subgiants,
which triggered an important discussion on the intermediate-age
populations in the outer GB.
We still need to search for microlensing events in order to obtain 
high-resolution spectra of dwarfs and subgiants
in regions with higher extinctions, including the NSD/CMZ,
which indicates the
importance of the infrared microlensing survey (Section~\ref{sec:Dwarfs}).
Considering the wide age distribution of the populations
in the inner regions, high-quality spectra obtained during
microlensing events would be useful for
investigating the history of star formation and chemical evolution
therein.
The detection of the He\,I {10,830\,{\AA}} line in GK giants will be
limited to the low-extinction region of the GB, $A_K \sim 0.5$~mag,
which would still provide important insights into
the stellar populations of the GB (Section~\ref{sec:He}).

\section*{Acknowledgements}

The author is grateful to colleagues,
in particular, A.~Dupree, S.~Hamano, N.~Kobayashi, P.~Whitelock, and T.~Yano, for valuable comments on the draft.
NM acknowledges the IAU grant supporting his attendance at
the IAU General Assembly and IAU Symposia.
He is supported by
a Grant-in-Aid (No.~18H01248)
from the Japan Society for the Promotion of Science (JSPS).

\begin{discussion}

\discuss{Valenti}{
Concerning the origin of the double red clump in the bulge,
\cite{Gonzalez-2015} presented that the scenario with
the helium-enhanced population (\cite[Lee \etal\  2015]{Lee-2015})
is unable to properly reproduce the RC luminosity distribution across
the whole bulge. Currently, there is no alternative explanation,
which is based on solid observational evidence,
to the X shape of the bulge.
}

\discuss{Matsunaga}{
I totally agree on the importance of considering
various observational results together to explain
the comprehensive properties of the bulge.
Supporting or disproving any of the interpretations
previously given for the issues discussed is
outside the scope of this contribution,
but in any case direct measurements of the helium abundances
would be very valuable for the related questions.
}
\end{discussion}

\end{document}